\documentclass[prl,twocolumn,showpacs,superscriptaddress,floatfix]{revtex4}
\usepackage{graphicx}

\begin{document}
\title{Breakdown of Diffusion in Dynamics of Extended Waves in Mesoscopic Media}

\author{A.A.~Chabanov}
\affiliation{Department of Physics, Queens College of the City
University of New York, Flushing, New York 11367, USA}

\author{Z.Q.~Zhang}
\affiliation{Department of Physics, Hong Kong University of
Science and Technology, Clear Water Bay, Kowloon, Hong Kong}

\author{A.Z.~Genack}
\affiliation{Department of Physics, Queens College of the City
University of New York, Flushing, New York 11367, USA}

\date{\today}

\begin{abstract}
We report the observation of nonexponential decay of pulsed microwave
transmission through quasi-one-dimensional random dielectric media that
signals the breakdown of the diffusion model of transport for
temporally coherent extended waves. The decay rate of transmission
falls nearly linearly in time due to a nearly gaussian
distribution of the coupling strengths of quasi-normal
electromagnetic modes to free space at the sample surfaces.
The peak and width of this distribution scale as $L^{-2.05}$ 
and $L^{-1.81}$, respectively.

\end{abstract}


\maketitle The diffusion model is widely applied to electronic,
neutron, and thermal conduction as well as to acoustic and
electromagnetic propagation in multiply scattering media. The
model is used not only when the phase of the wave is scrambled by
inelastic scattering, but also when the wave is temporally
coherent in mesoscopic samples \cite{ShengBook,Mesobook}. Although
wave interference leads to large intensity fluctuations within a
particular sample, the ensemble average of the flux reaching a
point is generally assumed to be the incoherent sum of
contributions of randomly phased, sinuating Feynman paths reaching
that point. In this model, the average intensity varies smoothly
in space and time and is governed by the diffusion equation.

Diffusion has been taken as the counterpoint to wave localization
\cite{Anderson,Gang4,Thouless77}. On one side are sharply defined
localized modes with the average level spacing $\Delta\nu$
exceeding the typical level width $\delta\nu$ and on the other 
are diffusing waves, for which $\delta\nu >\Delta\nu$ with
$\delta\nu\sim{D/L^2}$, where $D$ is the diffusion coefficient and
$L$ is the sample thickness \cite{Thouless77}. However, the mode
picture is inescapably a wave picture and at variance with
particle diffusion in a number of respects. First, the diffusion
equation is of first order, whereas the wave equation is of second
order in time. The time evolution of the wave at any instant should 
therefore depend not only upon the spatial distribution of the particle
density or intensity at that instant, as it does in the particle
diffusion picture, but also on the previous history of the wave.
Second, the particle picture represents the intensity as a
discrete sum over diffusion modes, whereas the wave picture
describes the field as a superposition of quasi-normal modes with
a continuum of decay rates in a random ensemble. The decay rates
of the diffusion modes are given by
$1/\tau_{n}=n^{2}\pi^{2}D/(L+2z_{0})^2$, where $n$ is a positive
integer and $z_{0}$ is the boundary extrapolation length. After a
time $\tau_{1}$, the intensity distribution settles into the
lowest diffusion mode and decays at a constant rate, $1/\tau_{1}$.
In contrast, the decay rates of quasi-normal modes in a random
ensemble should be a continuum. As time progresses, long-lived
quasimodes would contribute more substantially and the rate of
flow out of the sample would slow down continuously.
Nonexponential decay has been observed in acoustic scattering in
reverberant rooms \cite{BodlandKawakami} and solid blocks
\cite{Weaver} as well as in microwave scattering in cavities
whose underlying ray dynamics is chaotic \cite{Smilansky}.
Similarly, the decay rate of electronic conductance has been
predicted to fall as a result of the increasing weight of
long-lived, narrow-linewidth states \cite{Altshuler}. The leading
correction to the diffusion prediction for the electron survival
probability $ P_{s}(t)$ was calculated by Mirlin \cite{Mirlin00}
using the supersymmetry approach \cite{Efetov} to be, $-\ln
P_{s}(t) =(t/\tau_{1})(1-t/2\pi^{2}g\tau_{1})$, where $g$ is the
dimensionless conductance, which can be expressed as
$g={\delta\nu}/{\Delta\nu}$.

An ideal way to investigate the applicability of the diffusion
model to mesoscopic systems is to consider pulsed electromagnetic
transmission. Previous studies
\cite{McCall87,GenDrake89,Alfano90,Lagendijk97} have found
exponential decay for $t>\tau_{1}$, as predicted by diffusion
theory. However, measurements of optical transmission
\cite{Alfano90} indicate that the pulse rises earlier than
predicted by diffusion theory. Even more puzzling is the finding
by Kop \textit{et al.} \cite{Lagendijk97} of an increase in the 
inferred value of the diffusion coefficient with increasing $L$.

In this Letter, we report a dramatic breakdown of diffusion in
microwave measurements in nominally diffusive random samples for
which $\delta\nu>\Delta\nu$. We find that the decay rate of pulsed
transmission falls nearly linearly in time, as predicted by Mirlin
\cite{Mirlin00}. These results are interpreted in terms of the
distribution of decay rates of quasimodes of the sample, which is
found by taking the inverse Laplace transform of the decaying
signal. The distribution of the modal decay rates is nearly
gaussian, with an average value that scales as $L^{-2.05}$, which
is close to the inverse square scaling for the decay rate in the
diffusion model. The width falls as $L^{-1.81}$, which is faster
than predicted by Ref.~\cite{Mirlin00}.

Spectra of the in- and out-of-phase components of the steady-state
transmitted microwave field are measured in low-density
collections of dielectric spheres using a Hewlett-Packard 8772C
vector network analyzer. These spectra are multiplied by a
gaussian envelope of width $\Delta f$ centered at $f_c$ and then
Fourier-transformed to give the response to a gaussian pulse in
the time domain. Circular horns are positioned 30 cm in front of
and behind the sample. Linearly polarized microwave radiation is
launched from one horn and the cross-polarized component of the
transmitted field is detected with the other to eliminate the
ballistic component of radiation. The sample is composed of
alumina spheres of diameter 0.95 cm and refractive index 3.14,
contained within a 7.3-cm-diameter copper tube at an alumina
filling fraction of 0.068 \cite{Chabanov01}. This low density is
produced by embedding the alumina spheres in Styrofoam spheres of
diameter 1.9 cm and refractive index 1.04. Measurements for random
ensembles are obtained by momentarily rotating the tube about its
axis to create new random configurations before each spectrum is
taken. In this way, measurements are carried out in ensembles of
10,000 sample realizations at lengths of 61, 90, and 183 cm. In
addition, measurements are made in an ensemble of 2,300
realizations of a more strongly absorbing sample of 90 cm length
($L=90^{*}$ cm) produced by covering 40\% of the inside surface of
the tube with a thin strip of titanium foil laid from end to end.
The measurements are made within the frequency interval 14.7-15.7
GHz for $L=61$, 90 and 90$^{*}$ cm and 15.0-15.4 GHz for $L=183$
cm. The frequency intervals are chosen to be far from sphere
resonances \cite{Chabanov01}, so that the dynamics of transmission
is uniform over the frequency range and the sample is far from the
localization threshold. The closeness to localization, even in the
presence of absorption, is indicated by the variance of the
steady-state transmitted intensity normalized to its ensemble
average value, $var(I/\langle I\rangle)$ \cite{Nature}. In the
absence of absorption, $var(I/\langle I\rangle)\simeq 1+4/3g$
\cite{Nature,Variance}. At the localization threshold,
$var(I/\langle I\rangle)\simeq 7/3$. The values of $var(I/\langle
I\rangle)$ in the samples of $L=61$, 90$^{*}$, 90, and 183 cm are
1.18, 1.25, 1.26, and 1.50, respectively.
\begin{figure}[t!]
\includegraphics[width=\columnwidth]{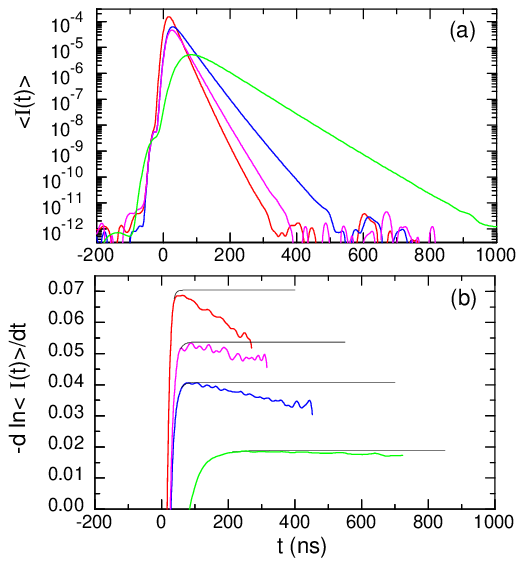}
\caption{(a) Average transmitted intensity in the alumina samples
of $L=61$ (red), $90^{*}$ (purple), $90$ (blue), and
$183$ cm (green). (b) Temporal derivative of the logarithm
of the intensity gives the rate $\gamma$ of the intensity decay
due to leakage out of the sample and absorption. The thin black 
curves are the decay rates of the diffusion model with $D^{*}$, 
$z_{0}$, and $\tau_{a}$ from Table~I.}
\end{figure}

The envelope of the temporal response to a gaussian pulse peaked
at $t=0$ is squared to give the transmitted intensity $I(t)$ for a
particular sample realization. The average transmitted intensity
$\langle I(t)\rangle$ is found by averaging over the ensemble, and
then over the frequency interval by shifting $f_{c}$. The results
are shown on a logarithmic scale in Fig.~1a. We find that, when
$\Delta f>\delta\nu$, the tail of $\langle I(t)\rangle$ does not
depend on $\Delta f$. We use $\Delta f=15$ MHz for $L=61$, 90, and
90$^{*}$ cm and $\Delta f=7.5$ MHz for $L=183$ cm, so that $\Delta
f>\delta\nu$ in all cases. The noise in the frequency spectra
produces a constant background intensity in the time domain. Once
this background is subtracted, a dynamic range of more than 6
orders of magnitude is achieved. This makes it possible to study
transmission on time scales an order of magnitude longer than the
time of peak transmission, though the longest times are still
smaller than the inverse level spacing, $1/\Delta\nu$, which gives
the time required for a photon to explore each coherence volume of
the sample. The measured decay rate due to leakage out of the
sample and absorption, $\gamma=-d\ln\langle I(t)\rangle /dt$, is
plotted in Fig.~1b. This rate is not constant as predicted by
diffusion theory, but falls nearly linearly with time. The constant 
increase in the decay rate for the $L=90^*$ sample over that for 
the $L=90$ sample, seen in Fig.~1b, indicates that the absorption rate
$1/\tau_{a}$ is constant in time. The decrease in $\gamma$ with
time is thus attributable solely to propagation out of the sample,
which proceeds at a rate, $\pi^{2}D(t)/(L+2z_{0})^{2}$. The
absorption rate is found from a fit of the diffusion model
\cite{Alfano90,McCall87,GenDrake89} to measurements of $\langle
I(t)\rangle$ up to the time at which 95\% of the full pulse energy
has been transmitted (Fig.~2).
\begin{figure}[t]
\includegraphics[width=\columnwidth]{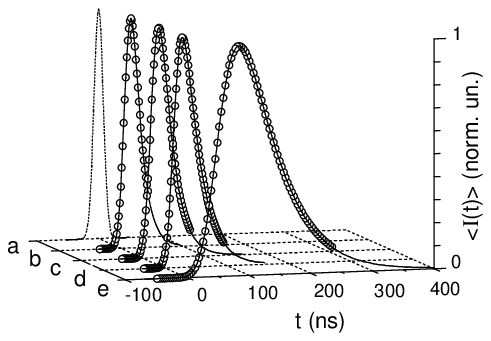}
\caption{Fit of the diffusion model to 95\% of the full pulse energy 
(circles) transmitted in the alumina samples of $L=61$ (b), $90^{*}$ (c), 
$90$ (d), and $183$ cm (e). The solid curves show the fit. 
(a) is the incident pulse of $\Delta f=15$ MHz used to obtain (b), (c), 
and (d); a pulse of $\Delta f=7.5$ MHz was used to obtain (e). 
All the curves are normalized.}
\end{figure}
The fit is obtained by taking $D(t)$ to be a constant, $D^{*}$,
and by minimizing the parameter $\chi^{2}=(\sigma^{2})^{-1}\sum[\langle
I(t)\rangle_{i}-I(t_{i})]^{2}$, where $\langle I(t)\rangle_{i}$
are the values of the measured intensity, $\sigma$ is the
uncertainty in $\langle I(t)\rangle_{i}$, averaged over the time
of the fit, and $I(t_{i})$ are the values of the model intensity
\cite{Alfano90} calculated at $t_{i}$. The parameters $D^{*}$, 
$\tau_{a}$, and $z_{0}$ obtained from the fit are listed in Table~I 
and the corresponding decay rates are shown by thin solid lines 
in Fig.~1b.
\begin{table}[b]
\caption{Values of the diffusion coefficient $D^{*}$, absorption
time $\tau_{a}$, and extrapolation length $z_{0}$, obtained from
fitting Eq.~(1) of \cite{Alfano90} to the short-time transmitted
intensity in Fig.~2.} \label{tab}
\begin{ruledtabular}
\begin{tabular}{ccrc}
$L$ (cm)   & $D^{*}$ (cm$^{2}$/ns) & $\tau_{a}$ (ns)  & $z_{0}$
(cm)  \\ \hline
 61         & 39.4 $\pm$ 0.3    &  [ 97 ]          & 9.6 $\pm$ 0.3 \\
  90           & 37.9 $\pm$ 0.3    & 104 $\pm$ 7      & 9.8 $\pm$ 0.6 \\
   90$^{*}$   & 37.4 $\pm$ 0.4    &  46 $\pm$ 2      &  8.7 $\pm$ 0.8 \\
    183        & 37.0 $\pm$ 0.8    &  97 $\pm$ 4      & 12.1 $\pm$ 2.5 \\
\end{tabular}
\end{ruledtabular}
\end{table}
For $L=61$ cm, $\chi^{2}$ at the minimum depends only weakly on
$\tau_{a}$ since the temporal range used in the fit is smaller than 
the $\tau_{a}$. For this reason, we use the value $\tau_{a}=97$ ns 
obtained in the fit to the data for $L=183$ cm for this length.

The diffusion coefficient $D^{*}$ obtained from the fit is found 
to decrease slightly with increasing $L$. The equality of the values 
of $D^{*}$ and $z_{0}$ for the samples with $L=90$ and $90^{*}$ cm, 
seen in Table~I, indicates that these parameters are not sensitive 
to the absorption rate. Moreover, when $1/\tau_{a}$ is subtracted 
from $\gamma$ to give the decay rates without absorption in Fig.~3a, 
the same time-dependent decay is found for both samples with length 
of $90$ cm.
\begin{figure}[t]
\includegraphics[width=\columnwidth]{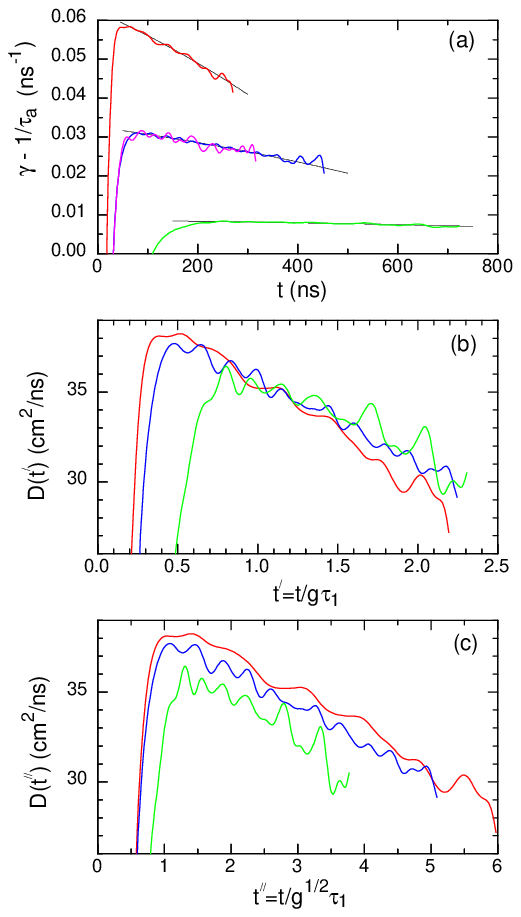}
\caption{(a) Leakage rates in alumina samples with $L=61$ (red),
$90^{*}$ (purple), $90$ (blue), and $183$ cm (green). The black
curves are the best fit to the data by a polynomial of power 2.
(b) ``Time-dependent diffusion coefficients'', $D(t)=(\gamma
-1/\tau_{a})(L + 2z_{0})^{2}/ \pi^{2}$, plotted versus $t'=t/g\tau_{1}$.
(c) $D(t)$ plotted versus $t''=t/\sqrt{g}\tau_{1}$.}
\end{figure}

The leakage rates in Fig.~3a give the ``time-dependent diffusion
coefficient'', $D(t)=(\gamma - 1/\tau_{a})(L + 2z_{0})^{2}/
\pi^{2}$. The values of $D(t)$ found from the measurements may be
compared to those from the theory of Ref.~\cite{Mirlin00} for the 
survival probability, which yields $D(t)=D(1-t/{\pi^{2}g\tau_{1}}+\ldots)$. 
This is done by plotting $D(t)$ as a function of the dimensionless 
time $t'=t/g\tau_{1}$ in Fig.~3b. The time $g\tau_{1}$ is proportional 
to $1/\Delta\nu$ and all the data are predicted to fall on a single 
curve. Although the data for samples with different values of $L$ at 
a given time appear to coincide within the noise, there is a clearly 
discernable decrease in the slope of $D(t')$ as $L$ increases. 
As a result, the curves do not extrapolate to a constant bare diffusion 
coefficient at $t=0$. When $D(t)$ is plotted instead versus the 
dimensionless time $t''=t/\sqrt{g}\tau_{1}$ in Fig.~3c, the slope 
of $D(t'')$ appears to be the same for all values of $L$, though 
the curves do not overlap. A strong deviation from exponential decay 
at $t''\approx 1$ has also been found in numerical simulations 
\cite{Casati97}. 

The changing slope of $D(t)$ with $L$ is associated with the scaling 
of the width of the distribution $P(\alpha)$ of decay rates of quasi-normal 
modes of the sample. These are hypothesized to form a complete set
\cite{Ching98}, even when $\delta\nu >\Delta\nu$. Since the time evolution 
is given by the superposition of these modes \cite{Ching98}, the average
transmission can be expressed as 
$\langle I(t)\rangle\propto\int_{0}^{\infty}P(\alpha)\exp(-\alpha t)d\alpha$, 
provided the mode coupling is, on the average, independent of $\alpha$. 
To find $P(\alpha)$, we use an approximate Laplace inversion algorithm 
based on the Weeks method \cite{InvLaplace}. A polynomial of power 2 is 
fit to the decay rates in Fig.~3a. These fits are then used to compute 
curves $\langle I(t)\rangle$, which are inverted to obtain the
distributions $P(\alpha)$ shown in Fig.~4. Note that a linear
decrease in $\gamma(t)$ with time as $\gamma(t)=a-bt$ would
correspond to a gaussian distribution of $P(\alpha)$ with
$\langle\alpha\rangle=a$ and $var(\alpha)=\sigma_{\alpha}^{2}=b$.
Since the decay of $\gamma(t)$ is nearly linear, $P(\alpha)$ is 
nearly gaussian.
\begin{figure}[t]
\includegraphics[width=\columnwidth]{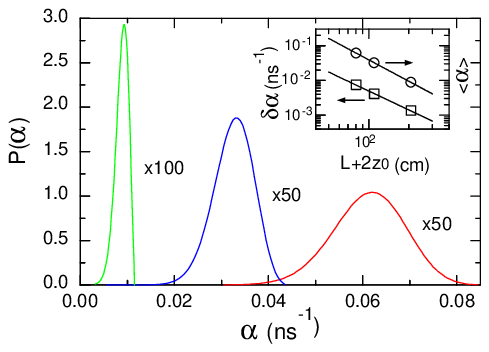}
\caption{Distribution of the modal decay rates, $P(\alpha)$, in
the alumina samples of $L=61$ (red), $90$ (blue), and
$183$ cm (green). Inset: scaling of the average
$\langle\alpha\rangle$ (circles) and of the width
$\sigma_{\alpha}$ (squares) of $P(\alpha)$, and the corresponding
best fit to $\sim (L+2z_{0})^{\beta}$ (solid) are shown on the
log-log scale. The best fits are obtained with exponents of
$-2.05$ and $-1.81$ for $\langle\alpha\rangle$ and $\delta\alpha$,
respectively.}
\end{figure}

The scaling of the average decay rate due to leakage out of the
sample, seen in Fig.~4, is given by $\langle \alpha \rangle
\propto (L+2z_{0})^{-2.05 }$. This is close to the inverse square
scaling of the diffusion model and suggests that the dynamics
observed is characteristic of extended waves and is not associated
with the approach to localization with increasing $L$. The width
of the distribution scales as $\sigma_{\alpha}\propto{(L+2z_{0})^
{-1.81}}$. This is close to the scaling,
$\sigma_{\alpha}\propto{\sqrt{b}}\propto 1/{g^{1/4}\tau_{1}}\propto
L^{-1.75}$, and differs from the scaling predicted by Ref.~\cite{Mirlin00}, 
$\sigma_{\alpha}\propto\sqrt{b}\propto 1/{\sqrt{g}\tau_{1}}\propto L^{-1.5}$. 
 
The wide distribution of the modal decay rates in thin samples may
be the source of the sharp spectral peaks observed in amplifying
random media \cite{Cao99,Frolov98}. In these samples, lasing would
occur in the longest-lived modes, which have the lowest critical
gain \cite{Sebbah01}.

In conclusion, we have found nonexponential decay of pulsed
transmission through disordered media in which the level width
exceeds the spacing between levels, even at long times and in
thick samples. This departure from diffusion theory is interpreted
in terms of the decay rate statistics of electromagnetic
quasi-normal modes. The statistics of these modes is fundamental to
understanding the static and dynamic behavior of waves in both
passive and active random media.

\begin{acknowledgments}
Discussions with X.~Zhang, R.L.~Weaver, T. Kottos, B.~Shapiro, and
B.A.~van~Tiggelen, are gratefully acknowledged. This research is
sponsored by the National Science Foundation (DMR0205186) and by
the U.S. Army Research Office (DAAD190010362).
\end{acknowledgments}

\end{document}